# RSC remodeling of oligo-nucleosomes: an atomic force microscopy study

**Fabien Montel[1,2], Martin Castelnovo[1,2], Hervé Menoni[3], Dimitar Angelov[3], Stefan Dimitrov[4]\* and Cendrine Faivre-Moskalenko[1,2]\***

[1]Université de Lyon, Laboratoire de Physique, CNRS UMR 5672, Ecole Normale Supérieure de Lyon, 46 allée d'Italie, 69364 Lyon Cedex 07, France ;

[2]Université de Lyon, Laboratoire Joliot-Curie, CNRS USR 3010, Ecole Normale Supérieure de Lyon, 46 allée d'Italie, 69364 Lyon Cedex 07, France ;

[3]Université de Lyon, Laboratoire de Biologie Moléculaire de la Cellule, CNRS UMR 5239, Ecole Normale Supérieure de Lyon, 46 allée d'Italie, 69364 Lyon Cedex 07, France ;

[4]INSERM, Université Joseph Fourier-Grenoble 1

Institut Albert Bonniot U823, Site Santé, BP 170, 38042 Grenoble Cedex 9, France.

\* **corresponding authors :** <u>Cendrine Faivre-Moskalenko</u> : Laboratoire de Physique, Ecole Normale Supérieure de Lyon, 46 Allée d'Italie, 69364 Lyon cedex 7, France. E-mail: <u>cendrine.moskalenko@ens-lyon.fr</u>, Tel : + 33 4 72 72 83 75, Fax : +33 4 72 72 89 50; and <u>Stefan Dimitrov</u> : Institut Albert Bonniot U823, Site Santé, BP 170, 38042 Grenoble Cedex 9, France. E-mail : <u>stefan.dimitrov@ujf-grenoble.fr</u>, Tel : + 33 4 76 54 94 73, Fax: +33 4 76 54 95 95.

**Keywords**: AFM, RSC, chromatin, remodeling, dinucleosomes,

**ABSTRACT** :

RSC is an essential chromatin remodeling factor that is required for the control of several processes including transcription, repair and replication. The ability of RSC to relocate centrally positioned mononucleosomes at the end of nucleosomal DNA is firmly established, but the data on RSC action on oligo-nucleosomal templates remains still scarce. By using Atomic Force Microscopy (AFM) imaging, we have quantitatively studied the RSC-induced mobilization of positioned di- and trinucleosomes as well as the directionality of mobilization on mononucleosomal template labeled at one end with streptavidin. AFM imaging showed only a limited set of distinct configurational states for the remodeling products. No stepwise or preferred directionality of the nucleosome motion was observed. Analysis of the corresponding reaction pathways allows deciphering the mechanistic features of RSC-induced nucleosome relocation. The final outcome of RSC remodeling of oligosome templates is the packing of the nucleosomes at the edge of the template, providing large stretches of DNA depleted of nucleosomes. This feature of RSC may be used by the cell to overcome the barrier imposed by the presence of nucleosomes.

**INTRODUCTION**

Chromatin is highly organized. The first level of chromatin organization, the nucleosome, consists of an octamer of core histones (two of each H2A, H2B, H3 and H4), around which ~ 165 bp of DNA is wrapped (1). The nucleosomes are connected by linker

DNA, which is usually associated with the linker histone. The presence of the linker histone facilitates the folding of the poly-nucleosomal filament into the 30 nm chromatin fiber.

Chromatin organization is highly dynamical and sensitive to small modifications like the replacement of a conventional histone by one of its variant (2-6) or the post-translational modification of the histone tails (7,8). The histone tails and their modifications are also essential for the organization of the mitotic chromosomes (9,10).

One of the main players in chromatin dynamics are the remodeling factors. These high molecular multi-protein complexes can reorganize and remodel the structure of chromatin at the expense of the energy freed by the hydrolysis of ATP. There are at least four different families of chromatin remodelers, namely the SWI2/SNF2, ISWI, CHD and INO80 families (11). A common property of all members of the different families is to mobilize nucleosomes (12). The general mechanism of nucleosome mobilization by chromatin remodelers remains, however, elusive (reviewed in (13-15)).

The yeast RSC (Remodels Structure of Chromatin) complex, a part of the SWI2/SNF2 family of chromatin remodelers (16) plays an essential role in the control of transcription, in the repair of damaged DNA and in the segregation of chromosomes (17-19). RSC, like SWI/SNF (another member of the SWI2/SNF2 family), has a centrally cavity, large enough to accommodate the binding of a single nucleosome (20,21). Recent electron cryo-microscopy data suggests that RSC is able to remodel only one nucleosome at the time (22).

*In vitro* experiments firmly established the ability of RSC and SWI/SNF to relocate centrally positioned mononucleosomes at the end of the nucleosomal DNA (2, 23-25). Interestingly, this observation sustains for long mononucleosomal template as recently shown using AFM and high resolution PAGE experiments (26). However the data on the action of RSC and SWI/SNF on oligo-nucleosomal templates remain scarce (27-29). Recent experiments of di- and tri-nucleosomes remodeling by SWI/SNF suggest that this particular remodeler has some nucleosome disassembly abilities (30).Other available works report evidence for a capacity of both remodelers to disorganize regularly spaced chromatin templates and to generate in some cases tightly packed nucleosomes (31-34).

Focusing on very detailed conformational analysis of mononucleosomes as deduced from the combination of high resolution AFM imaging and automated AFM image analysis, we have been able to demonstrate very subtle modifications of the structural and dynamical properties of the nucleosome due to the remodeling and sliding by SWI/SNF or the incorporation of histone variant (6,25). More recently the same kind of approach contributed to highlight the existence of a metastable intermediate state during RSC action on mononucleosomes, where the nucleosome is remodeled (change of complexed DNA length) but not slid (no change in position) (35). This intermediate, termed remosome, might have important functional implications (35).

In this work, by using Atomic Force microscopy (AFM), we have studied how RSC acts on di- and trinucleosomal templates, as well as oriented mononucleosomal templates bearing one end of the free DNA arms labeled with streptavidin. This last template allowed the analysis of the directionality of the RSC-induced nucleosome mobilization. Our data show for all templates that RSC packs nucleosomes at the end of the template, increasing the length of stretches of nucleosome-free DNA. Quantification of the results indicates the possible pathways of the mobilization reaction. Nucleosomes were mobilized by RSC equally well in both directions and no dependence of the efficiency of mobilization on the underlying DNA sequence was observed.

## MATERIALS AND METHODS

### Preparation of DNA probes

To prepare the nucleosomal templates we use the 601 positioning sequence (36). In order to produce orientated repetitions of the 601 sequence, two (or three) 601 DNA sequences were ligated using non-palindromic restriction site. This gives two DNA templates of 496 bp and 591 bp respectively for di- and trinucleosomes. The DNA fragments are cut from a plasmid containing either the di- or trinucleosome template and the DNA band of the desired size is cut and purified from agarose gel with a *promega kit.*

For the dinucleosomes, the DNA linker size is 58 bp and the DNA free arms are 63 bp and 81 bp long. For the trinucleosomes, the DNA linkers are 50 bp long, and the free DNA arms are 25 bp long on each side.

### Proteins purification and Nucleosome reconstitutions

Recombinant Xenopus laevis full-length histone proteins were produced in bacteria and purified as described (37).

Di- and trinucleosome reconstitutions were performed by the salt dialysis procedure (38). Purified DNA was mixed with equimolar amount of histone octamer in nucleosome reconstitution buffer NRB (2 M NaCl, 10 mM Tris pH 7.4, 1 mM EDTA, 5 mM β-mercaptoethanol). To find the initial range of concentration, we gel quantified histone and DNA solutions separately. Then, the optimum histone octamer to 601 DNA ratio was determined empirically by carrying out titrations over a molar ratio range 0.8–1.5 histone octamers per 200 bp 601 DNA repeat. The final dialysis step was performed in a buffer containing 10 mM Tris pH 7.4, 0.25 mM EDTA and 10 mM NaCl. Yeast RSC complex was prepared as described (16).

**Streptavidin labeled mononucleosomes**

The 5'-end of the 601 DNA template (of total length 356 or 311 bp) is biotin labeled using biotinylated primers during PCR amplification of the 601 DNA fragment from (plasmid pgem.). Biotinylated DNA is then added to histone mix for salt dialysis mononucleosome reconstitution. After reconstitution, the nucleosomes are incubated with a large excess of streptavidin proteins overnight, and the excess unbound streptavidin is then removed by Microcon filtering. In those binding conditions, we can check using AFM imaging, that all the nucleosomes (or free DNAs) are streptavidin labeled, but that only one nucleosome is linked to a streptavidin (4 biotin binding sites per streptavidin). The biotin-streptavidin link is strong enough ($K_d \sim 10^{-15}$ M) to ensure that the streptavidin will remain fixed during nucleosome sliding by RSC. Streptavidin labeled mononucleosome sliding and AFM imaging is performed as described for polynucleosomes

**Sliding experiment for AFM visualization**

The sliding of poly-nucleosomes for AFM experiments was performed by incubating di- or trinucleosomes (~ 20 ng/µl) with different concentrations of RSC (as indicated) at 29°C and in remodeling buffer containing 10 mM Tris-HCl (pH = 7.4), 1.5 mM $MgCl_2$ and 1 mM ATP. The reaction was stopped after the time indicated, by ~ 10 time dilution in TE buffer (Tris-HCl 10 mM, pH = 7.4, EDTA 1 mM) and deposition onto the functionalized APTES-mica surface.

**Atomic Force Microscopy and surface preparation**

For the AFM imaging, the dinucleosomes were immobilized onto APTES-mica surfaces. The functionalization of freshly cleaved mica disks (muscovite mica, grade V-1, SPI) was obtained by self-assembly of a monolayer of APTES under Argon atmosphere for 2 hours (39). A 5 μl droplet of the nucleosome solution is deposited onto the APTES-mica surface for 1 min, rinsed with 1 mL of milliQ-Ultrapure© water and gently dried by nitrogen flow. The samples are visualized using a Nanoscope III AFM (Digital Instruments™, Veeco, Santa Barbara, CA). The images have been obtained in Tapping Mode in air, using Diamond Like Carbon Spikes tips (resonant frequency ~150 kHz) at scanning rates of 3 Hz over scan areas of 1 μm wide.

**Image analysis**

AFM images were flattened with a simple height criterion and over-sampled two times. The proportions of each type of dinucleosomes were determined by manual counting over ~300 images which represent ~1000 dinucleosomes per experimental condition. In all cases several persons performed counting independently. Corresponding measurements were in a 5% range standard deviation. Similarly, the counting of trinucleosomes and streptavidin labelled mononucleosomes was manually performed.

**RESULTS**

**Biochemical characterization of the reconstituted dinucleosomal templates**

To study the RSC-induced different configurational states of the dinucleosome we have used a tandem repeat of the 601 DNA sequence. This has allowed the reconstitution of a well-defined dinucleosomal template, containing precisely positioned nucleosomes. The two individual nucleosomes within the dinucleosome were separated by 58 bp of linker DNA and contained two free DNA arms of 63 bp and 81 bp, respectively (see Figure 1 for details). The EMSA shows essentially no presence of free DNA in the reconstituted sample, indicating a complete association of the available dinucleosomal DNA with the histone octamers under our experimental conditions of reconstitution (Figure 1a, lane 1). Partial digestion with Dra III of the reconstituted dinucleosome bearing a Dra III restriction site in the middle of the linker DNA resulted in the generation of mononucleosomes only (no free DNA was detected, Figure 1a, lane 2), further demonstrating that the two 601 sequences were associated with two histone octamers. In agreement with this, AFM visualization shows the presence of two nucleosomes within the dinucleosomal template (Figure 1c.)

To characterize the structure of the reconstituted samples in more detail we have also carried out DNase I footprinting analysis for the reconstituted di- and control mononucleosomes as well as their respective naked DNA substrates (Figure 1b). The DNAse I footprinting of the dinucleosomes clearly shows two regions exhibiting strong protection, characteristic of well positioned nucleosomes within the dinucleosome, separated by a region showing a digestion pattern of naked DNA and corresponding to the linker DNA (Figure 1b, lane 1). The DNase I digestion pattern of the protected region is identical to that of the corresponding reconstituted control mononucleosome (Figure 1b, lane 3). Taken together, these biochemical data demonstrate that the reconstituted dinucleosomes are homogenous and precisely positioned. We have next used these well-characterized templates for the visualization by AFM of RSC mobilization experiments.

**RSC produces quantized dinucleosomal configurations**

The remodeling reaction on dinucleosomes was studied by taking snapshots of the dinucleosomes states at various times during the action of RSC. This was essentially carried out as described previously for the visualization of the sliding of mononucleosomes (25). Briefly, RSC and ATP were added to a solution of positioned dinucleosomes and incubated for one hour at 29°C. The remodeling reaction was stopped by dilution of the mix and deposition on AP-Mica for AFM imaging in air. A representative set of images is shown on Figure 2, where the control experiment is done with no ATP in the solution. RSC complex bound to nucleosome can be occasionally observed (typically less than 10% for low RSC concentration) and clearly identified as a larger object. These events were further disregarded from our image analysis in order to focus on the evolution of dinucleosomal states between successive steps of RSC action. AFM images show that incubation with RSC in the absence of ATP does not change the organization of the dinucleosomes, i.e. these particles were indistinguishable from the initial, non-incubated with RSC dinucleosomes (compare Figures 2a, 2c with Figure 1).

The picture is, however, quite different for the samples incubated in the presence of both RSC and ATP (Figure 2b and 2d). Indeed, new configurational states of the dinucleosomes were observed. A closer inspection of the AFM images shows that these new dinucleosomal configurational states were quantified. Remarkably, only the following five distinct configurational states were found (Figure 2e):

State #1: two nucleosomes positioned in the 601 positioning sequences (initial state).

State #2: one nucleosome is positioned in a 601 positioning sequence whereas the second nucleosome has been relocated at the DNA fragment end. Actually this state is more generally described by the first nucleosome located at the extreme end position, whereas the second nucleosome exhibited free DNA linker on each side.

State #3: two closely packed nucleosomes, generated upon the relocation of one nucleosome to the non-mobilized second nucleosome positioned in the 601 sequence.

State #4: the two nucleosomes have been relocated at the same end of the DNA fragment and are closely packed

State #5: each individual nucleosome has been relocated at a distinct end of the DNA fragment.

We would like to note that states #2, 3, 4 are degenerate in AFM images since both nucleosomes of the template are undistinguishable. We took this effect into account in our analysis (for example by gathering states #2 and #2bis within a single state #2). Note however that our conclusions are not dependent on this degeneracy.

In order to characterize in an alternative way the position of nucleosomes, one might be tempted to perform DNA contour length on AFM images, just like in one of our previous work (25). However, for dinucleosomal templates, the simultaneous measurement of nucleosome position and DNA complexed length is impeded by the large inherent nucleosome fluctuations in addition to the mixing of 5 degenerate states. We checked indeed that this imply large uncertainties for the accurate localization of nucleosomes within a heterogeneous population (data not shown). Therefore, such a quantitative AFM image analysis is rather inefficient for our purpose of identifying global rules of RSC-induced nucleosome motion. Our semi-quantitative method represents the best compromise between spatial resolution and statistical sampling of dinucleosomal states.

Different RSC concentrations were used to perform the same type of experiment at fixed time of reaction ($t_0$ = 60 min) and ATP concentration (1 mM), where the RSC to nucleosome ratio is more than 10 times varied. For each condition 1000 dinucleosomes on average were analyzed (Figure 3). In all conditions tested, only the five above described configurational states of the dinucleosome were observed again, although their relative proportions depend on the concentration of RSC used in the mobilization reaction. As seen (Figure 3a), the proportion of the dinucleosomes in the initial state #1 decreases from 1 to 0 as the concentration of RSC increases. The two other states (#2 and #3) start to increase upon increasing the RSC concentration and then they gradually decrease reaching zero for the very last experimental condition (at the highest concentration of RSC). The states #4 and #5 increase gradually with RSC concentrations. For convenience, the RSC generated dinucleosome configurations could be divided, into three distinct groups. The first group (I) corresponds to the initial state #1 where no nucleosome has been mobilized. The second group (II) consists of the intermediate states #2 and #3 that represent, in a first approximation, the cases where only one of the two nucleosomes has been moved out of the 601 positioning sequence (towards either the other nucleosome or the end of the DNA fragment). Finally, states #4 and #5 constitute the third group (III), where both nucleosomes of the dinucleosome have been moved by RSC. The dependences of the dinucleosome proportions in each of these three distinct groups as a function of normalized RSC concentration are plotted in Figure 3b.

The same experiments were performed for a fixed motor (RSC) concentration ($v_0 = 0.4$ µl) and varying the time of incubation. By normalizing the time of reaction by $t_0$ and the motor concentration by $v_0$, the data superimpose (see Figure 3b) suggesting that time and motor concentration are equivalent control parameters towards the reaction coordinate under experimental time scale and concentration range used in this work. Such equivalence is consistent with the observed kinetics (Michaelis Menten kinetics) of most enzymes (40).

Noteworthy, RSC also generated products that contained only one nucleosome, i.e. one of the histone octamers from the dinucleosomal template was evicted in the reaction. The amount of these templates increases as a result of the remodeling reaction but this increase was relatively small and it was not taken further into consideration. (see supplemental Fig. S1).

**RSC induced nucleosome sliding is stopped by either neighbouring nucleosome or DNA end**

In order to understand quantitatively the evolution of population for each of the configurational states and to investigate further the mechanism of RSC action, we developed a simple analysis of the reaction pathway. First, the observation of a small limited number of discrete states strongly suggests simple rules for elementary move of single nucleosome induced by RSC. Indeed, the five states are characterized by only three typical nucleosome positions: either the nucleosome is still bound to its original 601 positioning sequence, or the nucleosome has reached one of the ends of DNA template; or finally the nucleosome is in a close vicinity to the other nucleosome located one of the two previous positions. If no intermediate states are observed, it implies that the probability of RSC dissociating from the substrate is low during translocation, but goes up significantly when an obstacle impeding further movement is reached. This simple observation suggests *a priori* that RSC-induced nucleosome motion in such templates is only stopped by physical barriers like a strong positioning sequence, a DNA end, or another nucleosome. The second critical observation is that the only surviving states for large reaction coordinate (either the time or the RSC concentration) are the states #4 and #5, while all the others states (#1,#2 and #3) decay toward zero population. Since in these two final states none of the nucleosome is located in the 601 positioning sequence, we conclude that, under saturating conditions, RSC-induced nucleosome motion is not significantly affected by underlying sequence. This imposes a more stringent restriction on the elementary nucleosome motion induced by RSC: such a motion is stopped either by a DNA end, or by another nucleosome.

As it is shown in the next section, this elementary rule for nucleosome relocation under RSC remodeling will also explain qualitatively the observations performed on tri-nucleosomes.

Using the previous observations, one can restrict the reaction pathway among the five states, like it is depicted on figure 4. This set of particular pathways is compatible with the observed irreversible evolution of the states towards a *quasi*-equilibrium between state #4 and #5. In particular, for large value of the reaction coordinate, the population of state #4 and #5 seem to equilibrate. Transitions from state #4 to state #5 and *vice-versa* are associated with a single nucleosome travelling from one end of the DNA to the opposite position next to the other nucleosome. This observation sets the bound of the maximal distance travelled by a single nucleosome mobilized by RSC to roughly 200 bp. Moreover, the larger population of state #5 near the *quasi*-equilibrium as compared to the population of state #4 indicates that end-positioned nucleosomes are less efficiently mobilized than nucleosomes neighbouring these end positions.

**RSC packs the nucleosomes at the edge of a tri-nucleosomal template**

By using dinucleosomal templates, we identified in the previous section a simple rule for the elementary nucleosome motion induced by RSC: such a nucleosome moves processively out of its original position until it encounters either another nucleosome or the end of the DNA template. In order to test this simple dynamical scheme for other nucleosomal template, we performed similar AFM measurements of RSC-sliding on a distinct template: a trinucleosomal template constructed by juxtaposition of three 601 positioning sequences.

The presence of the additional nucleosome in the case of trinucleosomal template increased the number of resulting reaction products. However the extension of the previous analysis to the case of trinucleosomes predicts the *quasi*-equilibrium configuration to be composed of only two distinct states: one where one nucleosome is located on one end and two others are located on the opposite end of the DNA (state #A) and the other where the three nucleosomes are located on the same end of the DNA template (state #B). Experimentally most of the trinucleosomes are indeed observed in state #A or #B for large reaction coordinates (Figure 5a). This strengthens the simple rule we have deduced for the elementary nucleosome motion induced by RSC. Interestingly, sequence effects did not appear to influence the outcome of RSC remodeling on trinucleosomal template since no nucleosome is localized within the strong 601 positioning signal in the final configuration. Moreover, the observed proportions of state #A and #B are respectively ~ 70% and 30% for a total of 174 trinucleosomes counted (Fig. 5b). This shows that end-positioned nucleosomes are less efficiently mobilized, in agreement with the similar observation made on dinucleosomal template.

**RSC mobilizes the nucleosomes in both directions on an oriented mononucleosomal template**

In order to further test both the reduced efficiency of RSC to mobilize end-positioned nucleosomes and the isotropy of mobilization, we performed a RSC sliding experiment on a streptavidin-end labeled mononucleosomal 601 template (Figure 6a). With this construction, it is possible to focus directly on the motion of one nucleosome in between one DNA end and a fixed obstacle mimicking a fixed nucleosome. The different size of a nucleosome compared to a streptavidin enables to identify the particle observed by AFM (see Figure 6a). For large reaction coordinate, we observe two *quasi*-equilibrium configurational states: in the first state #α, the nucleosome is located at the other DNA end opposite to streptavidin, while in the second state #β the nucleosome is located next to the streptavidin. The population of state #α is larger (~2/3 of the total) than the population of state #β (~1/3) for large reaction coordinate (Fig.6b). This provides another compelling evidence of reduced RSC mobilization efficiency of end-positioned nucleosomes.

In addition, changing the location of the end positioned streptavidin with respect to the DNA template (either at one end or at the other), has allowed us to address the question of directionality of mobilization of the nucleosome by RSC. The measured proportion of states #α and #β was found independent of the template construction for the three different constructions used : 127-601-82(s), 127(s)-601-82 and 82(s)-601-82 (where (s) indicates the end positioned streptavidin and the number report the length in bp of naked DNA arm on each side of the 601 positioning sequence). This means that the direction of sliding is independent of sequence or linker length asymmetry. We conclude that RSC induced motion is isotropic, which in turn implies that final states #α and #β are in equilibrium.

**DISCUSSION**

In this work we have studied the RSC remodeling of oligonucleosomal templates. The basic results were obtained by thoroughly analyzing the different conformational states of the RSC mobilized dinucleosomes. We have shown that RSC induces a limited set of configurational states and tends to pack nucleosomes at the edge of the dinucleosomal template. Since no intermediate localization of the mobilized nucleosomes was observed, this result strongly suggests the following simple rule for elementary nucleosome motion induced by RSC: a remodeled nucleosome moves processively out of its original position until it encounters either another nucleosome or the end of the naked DNA arm and then stops. This rule was further demonstrated to operate for "oriented" mononucleosomes (containing a bound streptavidin at one end of the free DNA arm) as well as for trinucleosomes. In addition, this observation of RSC-induced nucleosome motion arrest by physical obstacles is strengthened by the observation of RSC sliding for circular streptavidin labeled mono-nucleosomes (see supplemental Fig.S2). Indeed, in this case and for large reaction coordinate, all the mononucleosomes accumulate close to the streptavidin, the protein being the only physical obstacle in this construction (no DNA free end).

We have also found that the end-positioned nucleosomes were less efficiently mobilized by RSC, as compared to nucleosome positioned at any other site of the templates. Note that generally, the end-positioned nucleosome is referred as the final state of the RSC mobilization action (23). Our results illustrate that an end-positioned nucleosome can be still mobilized by RSC (albeit with a reduced efficiency), leading to *quasi*-equilibrium between 2 end-positioned nucleosomal states. The low efficiency of end-positioned nucleosome mobilization by RSC could reflect its non-canonical structure: indeed it has been demonstrated that remodeled nucleosome can be located slightly off the end (being therefore slightly under-complexed with less than 146 bp) (41,42).

In addition, as the ratio of state #$\alpha$ and #$\beta$ population is similar to the ratio of state #5 and #4 for dinucleosomes, it suggests that RSC is able to mobilize a positioned nucleosome indifferently toward another nucleosome or toward a streptavidin. Hence, it demonstrates that the reduced mobilization of end positioned nucleosome is not sensitive to size or charge effect of the physical barrier.

Our data also show that the underlying DNA sequence has no effect on the efficiency of the RSC induced mobilization of either di- or tri-nucleosomes or oriented mono-nucleosomes. In addition, no preferred directionality of nucleosome mobilization or step-wise mobilization was detected under our experimental conditions. These results differed from the recent experiments of RSC-induced mobilization of positioned mono-nucleosome on a long DNA template (26), where the authors noticed that nucleosome mobilization by RSC was more efficient when the nucleosome is not anymore under the influence of the initial positioning sequence. We cannot explain the origin of this difference for the moment. Noteworthy, the authors (26) found the majority of nucleosomes packed at the edge of DNA template for large reaction coordinate (time or RSC concentration), in agreement with the observations of the present work.

The simple rule for RSC-induced nucleosome motion identified in the present study can be used to infer the typical result of RSC action on longer poly-nucleosomal templates ($N$ nucleosomes): packing of nucleosomes on one side of the fragment or on both sides at the end of the DNA template is expected, freeing longer stretches of DNA from the presence of nucleosomes (~ ($N$+1) DNA linker size). Thus, a depletion of nucleosomes should be observed localized next to a nucleosome cluster accumulated due to the presence of a physical barrier (in our case the end of the DNA fragment). *In vivo*, such process might occur next to a barrier that could be a non-mobilizable variant nucleosome (H2A.Bbd for example (43)) or the presence of a transcription factor bound to the DNA (44,45). The RSC-induced nucleosome 'depletion' on oligomeric chromatin provides an appealing feature as one of the role of RSC in structuring and modulation of chromatin.

**FUNDINGS**

This work was supported by grants from INSERM, CNRS, the Association pour la Recherche sur le Cancer [Grant 4821], the Région Rhône-Alpes (Contrat Plan-Etat Région "Nouvelles Approches Physiques des Sciences du Vivant" and Convention CIBLE 2008), the Agence Nationale de la Recherche [ANR "EPIVAR" N° 08-BLAN-0320-02 to S.D and ANR CROREMBER NT09_485720 to D.A. and S.D.]. S.D. acknowledges La Ligue Nationale contre le Cancer (Equipe 19 labellisée La Ligue).

**ACKNOWLEDGMENTS**

We thank Dr. J. Workman for kindly providing us with the yeast strain expressing tagged RSC, and V. Gerson for isolation and purification of RSC.

**FIGURE LEGENDS**

**Figure 1: Characterization of the reconstituted positioned 601 dinucleosomes by EMSA, DNase I footprinting and AFM imaging.**

(a) 4% PAGE analysis of the full length 601 dinucleosomes (lane 1) and after partial digestion with DraIII (lane 2) and control mononucleosomes (lane 3).

(b) DNase I footprinting analysis of the di- (lane 1) and mono- (lane 3) nucleosomes. The respective naked DNA digestion profiles are also shown (lanes 2 & 4).

(c) AFM image gallery of reconstituted dinucleosomes deposited on AP-mica surfaces and imaged in tapping mode in air.

**Figure 2: AFM visualization of dinucleosome sliding by RSC.**

AFM images for dinucleosomes incubated for 60 minutes @ 29°C with the remodeling complex RSC and (a) in the absence of ATP and (b) in the presence of ATP. Typical AFM visualizations of (c) positioned dinucleosomes in the absence of ATP and (d) the 5 dinucleosome states observed in the presence of RSC and ATP (see text). (e) Schematic of the various dinucleosome states identified on the AFM images. The dinucleosomes have been reconstituted on two 601 positioning sequences separated by 58 bp DNA linker and with 63 bp on one side and 81 bp on the other side. States #1 to #5 correspond to the 5 states illustrated on the AFM images of (d).

**Figure 3: Experimental dinucleosome populations in each state: titration and time evolution.**

(a) Evolution of the proportion of dinucleosomes in each state (1 to 5) during the sliding reaction by RSC. Nucleosome populations are counted from analysis of AFM images obtained for various RSC/nucleosome ratio and a fixed incubation time of 60 minutes @ 29°C (the RSC/nucleosome ratio is estimated of ~1/40 for v = 0.4 µl). (b) The proportion of dinucleosomes in each group of states (I initial, II intermediate and III final) is reported as a function of general reaction coordinates: RSC volume (circles and dashed lines) or time (stars and solid lines). The x-axis has been normalized by $v_0 = 0.4$ µl of RSC for the titration and $\tau = 60$ min for the kinetics.

For each experimental data points, about 1000 dinucleosomes have been analyzed. Each state (#1 to #5) is illustrated by a typical AFM image of a dinucleosome in this state as defined on Fig.2, and the groups are defined as initial (I corresponding to state #1), intermediate (II corresponding to states #2 and #3) and final III (corresponding to states #4 and #5).

**Figure 4: Kinetic scheme showing the various states produced by RSC sliding of dinucleosomes.**

The states #1 to #5 have been constructed based on the quantized dinucleosomal states observed on the AFM images. The groups I to III are indicated as defined from the experimentally observed states #1 to #5, and correspond to initial (group I = state #1), intermediate (group II = state #2 and #3) and final (group III = state #4 and #5) state groups. The transitions arise from the simple rule for elementary nucleosome motion induced by RSC on a positioned di-nucleosome: a remodeled nucleosome moves processively out of its original position until it encounters either another nucleosome or the end of the naked DNA arm and then stop. Circular transitions correspond to iterations where the dinucleosome state is unchanged as a result of RSC action. Note that state #2 and #2bis are experimentally indistinguishable. The positions of the 601 positioning sequences are indicated in red, but are not to scale.

**Figure 5: RSC sliding of trinucleosomes**

(a) AFM images of trinucleosomes before and after the incubation with RSC and ATP.

(b) Schematic of the trinucleosome construction used. The DNA template contains three 601 positioning sequences separated by 50 bp DNA linker and with 25 bp on each side. The total length is 591 bp of DNA. Schematic of two main trinucleosome configurations identified on the AFM images for a strong sliding condition (high RSC concentration and large reaction time) corresponding to the stationary state. The observed proportion of trinucleosomes in each final state (#A and #B) is reported for a total number of N= 174 trinucleosomes counted.

**Figure 6: RSC sliding of streptavidin labeled mono-nucleosomes**

(a) Typical AFM images of streptavidin labeled mononucleosomes before and after the action of RSC and ATP for 3 different template constructions 127-601-82(s), 127(s)-601-82 and 82(s)-601-82 where (s) points for the end positioned streptavidin, and various naked DNA arm length (in bp) on each side of the 601 positioning sequence.

(b) Proportion of streptavidin labeled mononucleosomes mobilized by RSC away from the streptavidin (state #α) or towards the streptavidin (state #β) for 3 different template constructions. The total number of streptavidin labeled mononucleosomes counted was N = 854 for 127-601-82(s), N = 551 for 127(s)-601-82 and N = 447 for 82(s)-601-82.

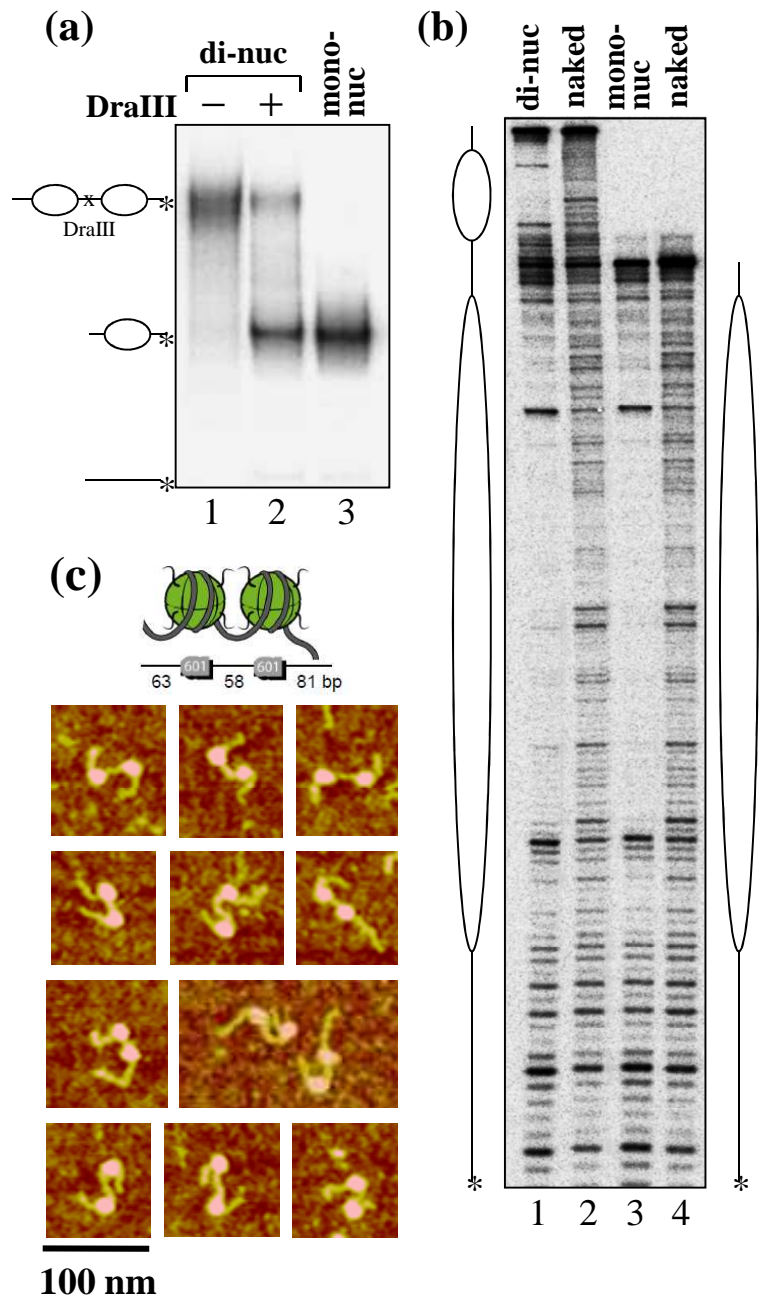

Figure 1_ Montel et al. 2010

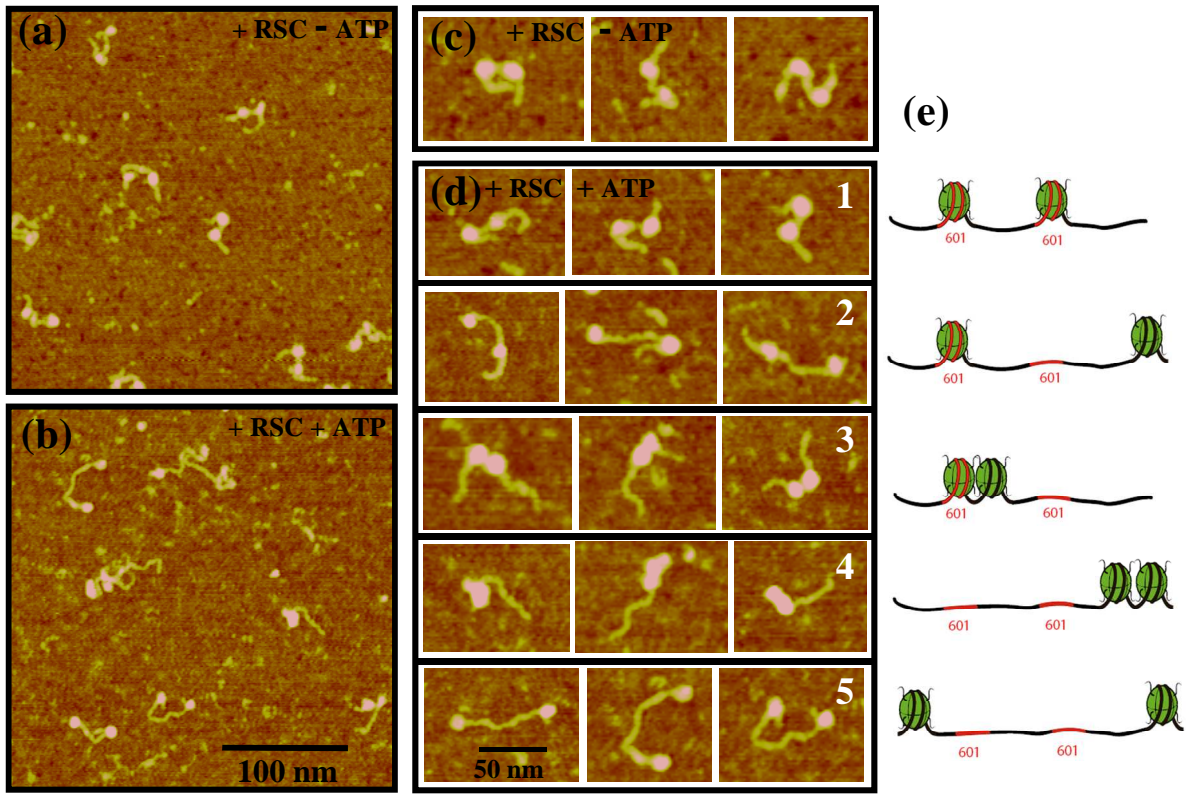

Figure 2_ Montel et al. 2010

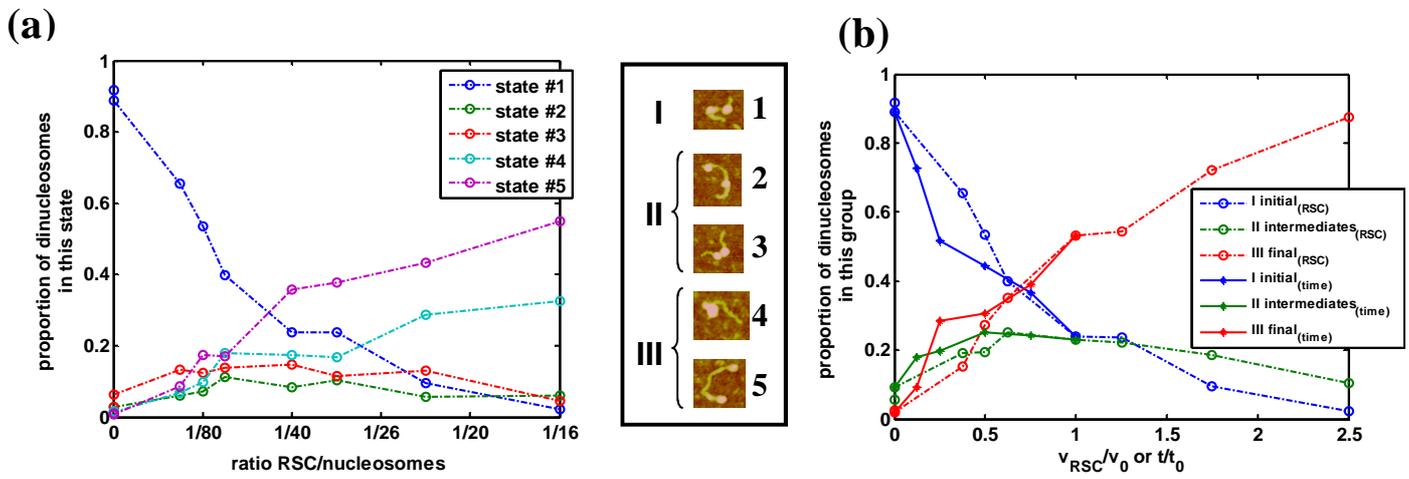

Figure 3_ Montel et al. 2010

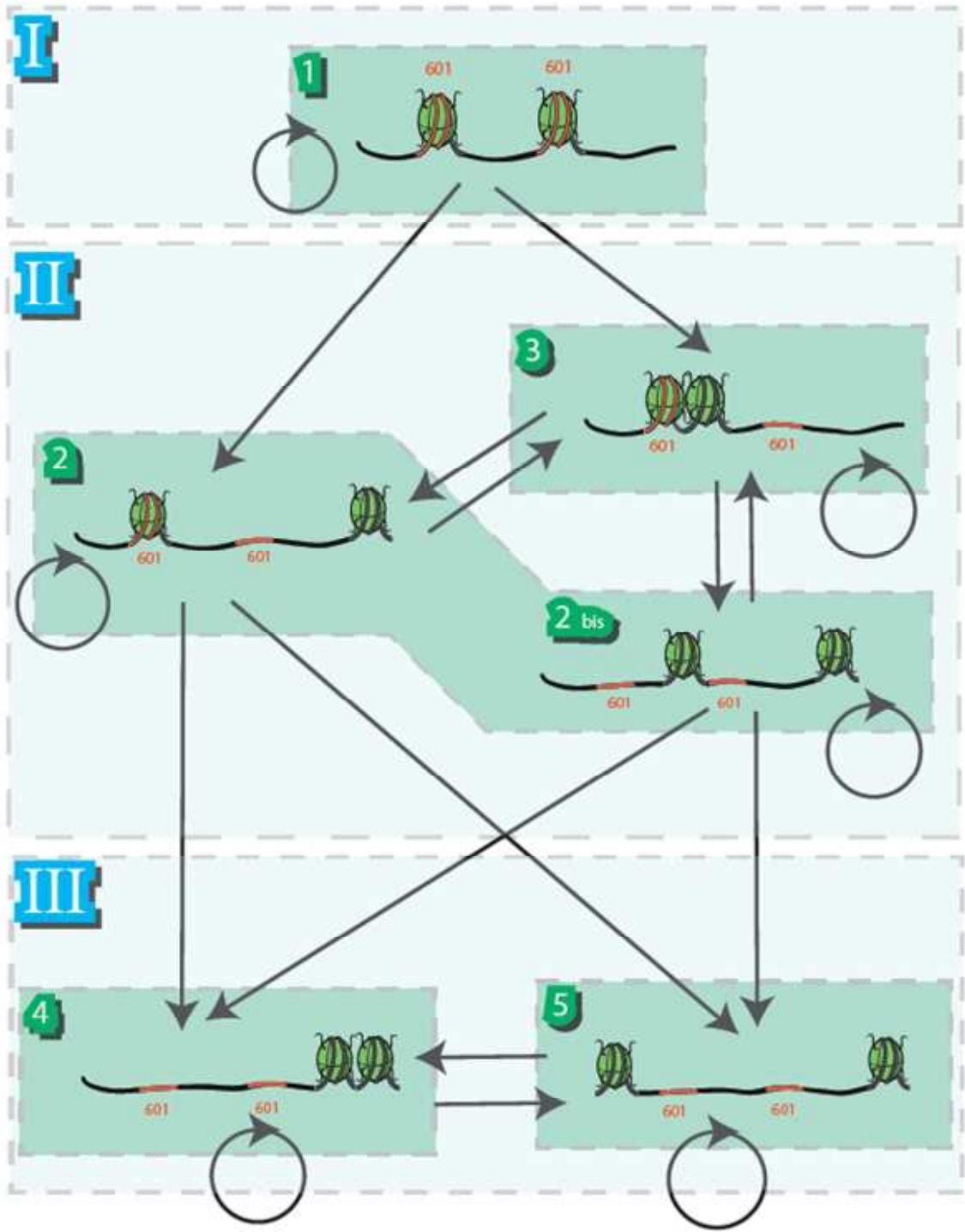

Figure 4_ Montel et al. 2010

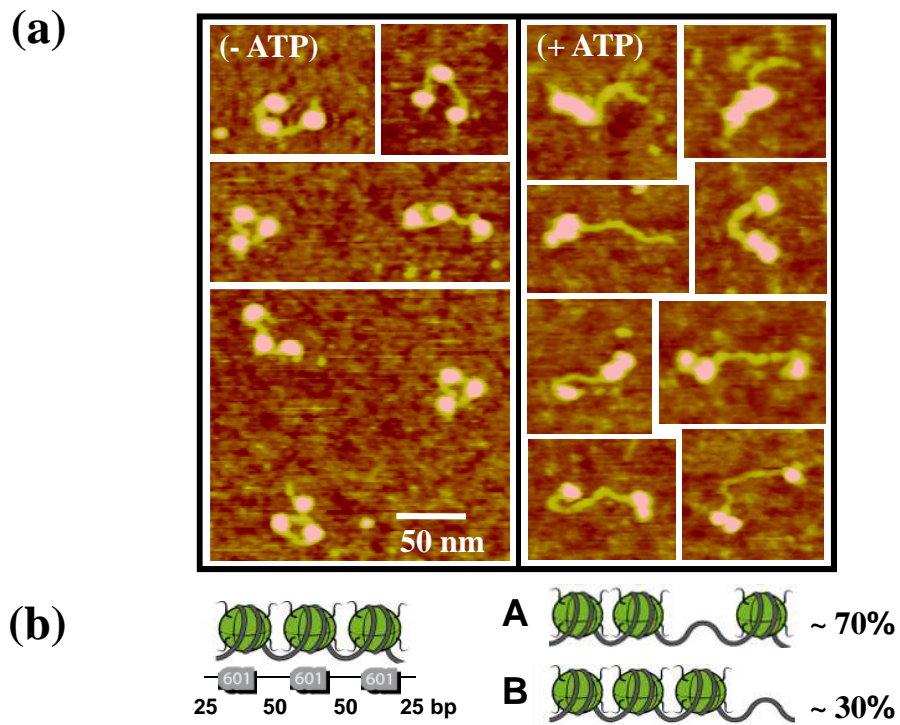

Figure 5_ Montel et al. 2010

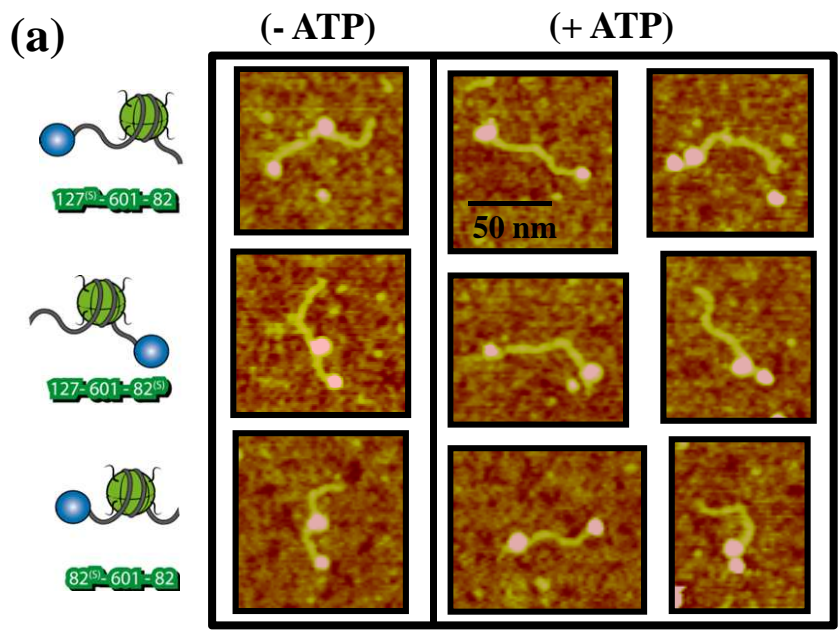

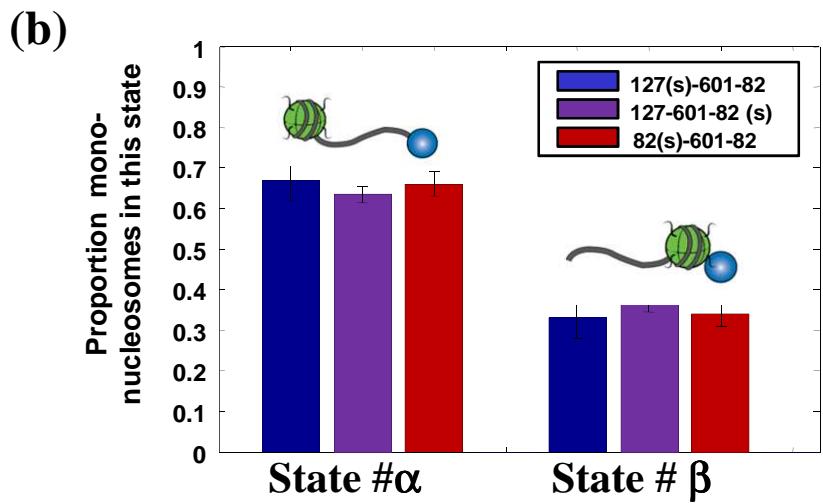

Figure 6_ Montel et al. 2010

# RSC remodeling of oligo-nucleosomes: an atomic force microscopy study

Fabien Montel, Martin Castelnovo, Hervé Menoni, Dimitar Angelov, Stefan Dimitrov and Cendrine Faivre-Moskalenko

## Supplemental Material

### 1- Few mononucleosomes are induced by RSC action on dinucleosomal templates

We show here the sliding reaction of dinucleosomes as a function of RSC volume, including the counting of mononucleosomes during RSC mobilization of dinucleosomes (Fig. S1a). Mononucleosomes are observed only in 2 positions: either within 601 position or at the end of DNA template. For each condition, the mononucleosome proportion corresponds to the total number of mono-nucleosomes. Fig. S2 displays the detailed composition of mononucleosome population throughout the sliding reaction of Fig.S1a.

For zero or low RSC/nucleosome ratio, only 601 positioned mononucleosomes are found, that correspond to slightly undersaturated dinucleosome reconstitution. This number is less than 10% of the total number of nucleosomal template at room temperature. Note that for large reaction coordinate, most of the mono-nucleosomes are end-positioned.

We observe mainly two features: (i) the proportion of mononucleosomes does not increase as soon as RSC produces new states (#2 to #5), but only when states #4 and #5 become significant.

(ii) the rate of new mononucleosome appearance seems to follow the rate of production of state #5. This suggests that new mononucleosomes are mainly produced through the RSC action on state #5 nucleosomes by ejection of one end-positioned nucleosome of the di-nucleosomal template.

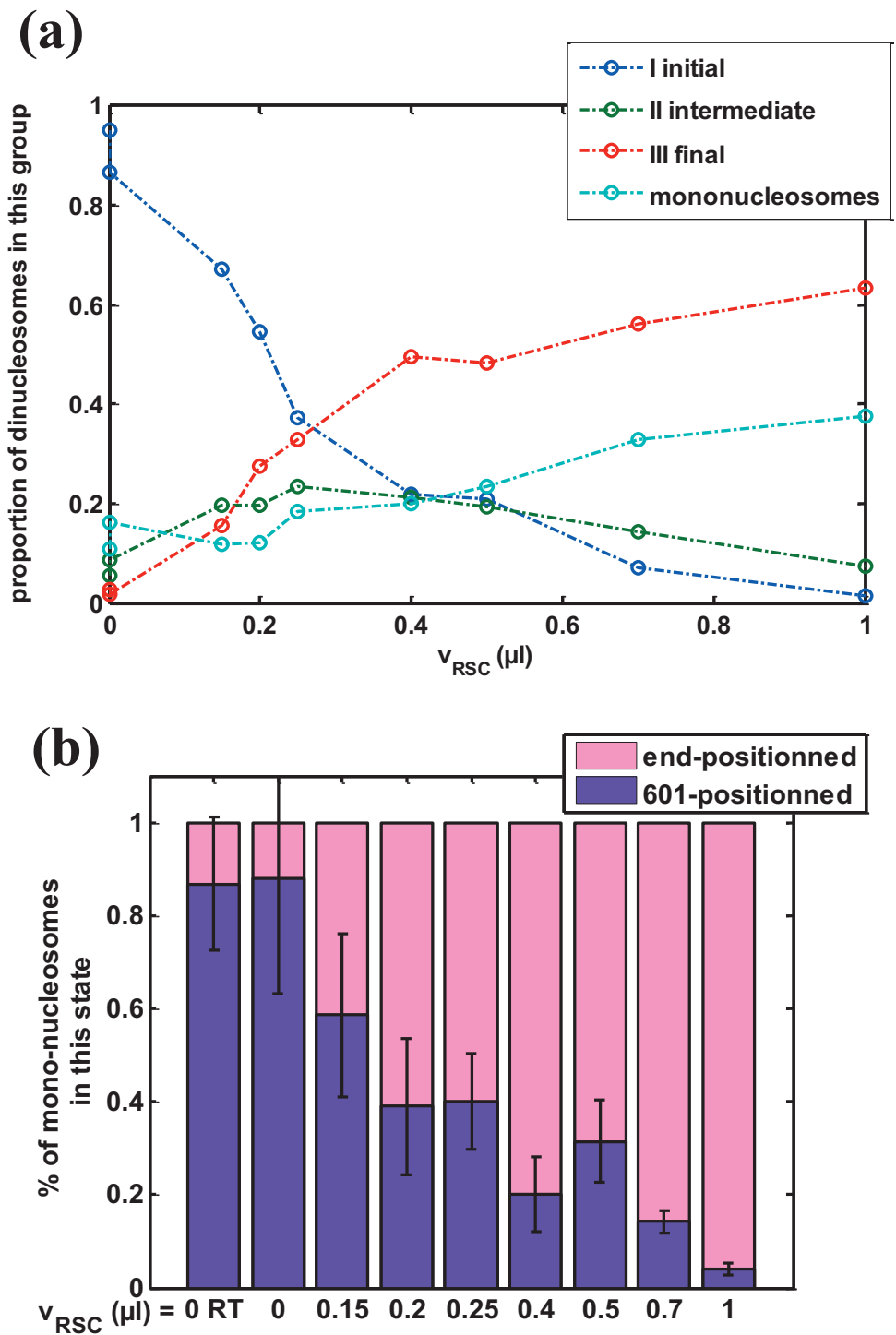

**Figure S1 : Population distribution of di- and mononucleosomes during RSC action.**
(a) The proportion of dinucleosomes in each group of states (I initial, II intermediate and III final) as well as the proportion of mononucleosomes is reported as a function of RSC volume. (b) For mononucleosomes, the total amount shown on fig.S1a is decomposed into the proportion that is 601-positioned and the other part that is end-slided.
Note that $v_{RSC} = 0.4\mu l$ corresponds to a 1/40 RSC to nucleosome ratio.

## 2- RSC does not need a free DNA end to slide mononucleosomes.

To complement the results presented in the manuscript on streptavidin labeled mononucleosomes sliding by RSC, we performed experiments with circular templates by adding a new biotin tag to the other end of DNA (compared with the previous situation with only one biotin tag at an end and therefore one single binding site for the streptavidin.).

With this double-tags construction and for low streptavidin concentrations, the major conformation of reconstituted nucleosomes is circular one (Fig. S2a), one streptavidin being attached to both ends of the DNA. We performed RSC remodeling reaction on this nucleosomal substrate, and the results are presented in supplemental Figure S2. For linear streptavidin labeled templates, 93% of the nucleosomes are slided in the +RSC condition. We would like to note that this population is actually composed of 65% mononucleosomes slided away from the streptavidin (state #$\alpha$) and 35 % mononucleosomes slided against the streptavidin (state #$\beta$) as it is shown in fig. 6b of the manuscript. For circular templates, there is a single slided position that is against the streptavidin (the template is not anymore oriented), and 88% of the mononucleosomes in the +RSC condition are in this state. The main conclusion about these new experiments is that RSC does not need any DNA end in order to mobilize nucleosomes, showing similar efficiency for both linear and circular templates in the same sliding conditions (same RSC/nucleosome ratio, ATP concentration, Temperature and incubation time, etc).

The high sliding efficiency observed for circular template also strengthens the message of our work about the arrest of RSC-induced nucleosome motion by physical obstacles: indeed the only physical obstacle in this construction being able to stop the RSC-induced motion is the streptavidin.

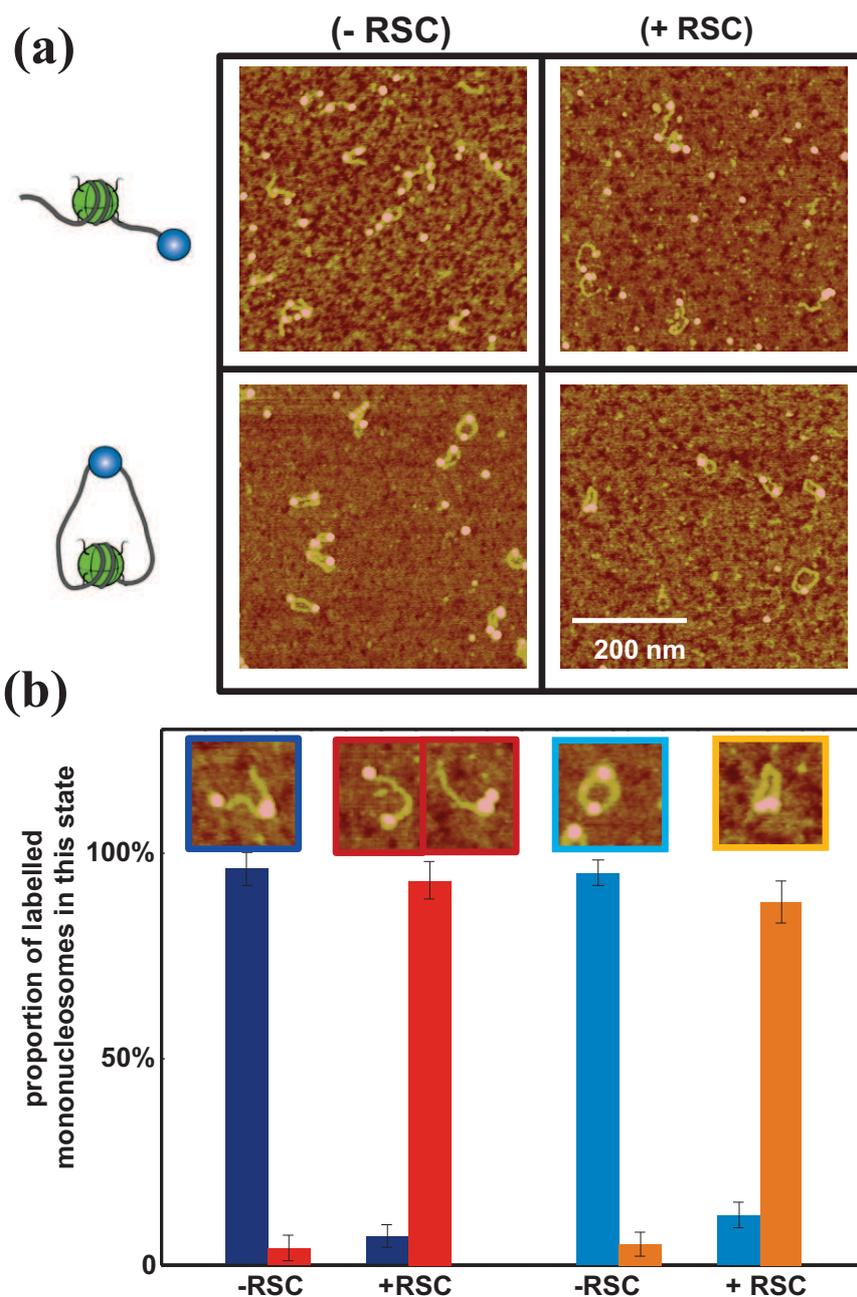

**Figure S2 : Comparison of RSC sliding efficiency for linear (single biotin tag) or circular (double biotin tag) streptavidin labeled mononucleosome .**
(a) typical AFM images of single tag linear (top) and double tag circular (bottom) streptavidin labeled mononucleosomes without (left) or with (right) RSC.
(b) Counting of streptavidin labeled mononucleosomes in each state: linear-601-positioned (dark blue), linear-end-positioned (red), circular-601-positioned (light blue), circular-end-positioned (orange). Typical AFM image of each stat is shown with the corresponding colored frame. The number of mono-nucleosomes analyzed in this experiment is: N(-RSC) = 468 and N(+RSC) = 766 for linear streptavidin labeled mononucleosomes, N(-RSC) = 499 and N(+RSC) = 277 for circular streptavidin labeled nucleosomes.

*Supplemental methods: Both ends of the 601 DNA template (of total length 356 or 311 bp) is biotin labeled using 5'-biotinylated primers during PCR amplification of the 601 DNA*

*fragment from (plasmid pgem.). Biotinylated DNA is then added to histone mix for salt dialysis mononucleosome reconstitution. Streptavidin labeling, mononucleosome sliding and AFM imaging is performed as described in the Material and Methods section of the manuscript.*